\documentclass[journal]{IEEEtran}
\usepackage[edges]{forest}
\usetikzlibrary{arrows.meta}
\usepackage{graphicx,times,amsmath,cite,enumerate}
\usepackage{kantlipsum}
\usepackage{threeparttable}
\usepackage{lipsum}
\usepackage{epsfig}
\usepackage{float}
\usepackage{amssymb}
\usepackage{latexsym}
\usepackage{siunitx}
\usepackage{psfrag}
\usepackage{setspace}
\usepackage{color}
\usepackage{verbatim}
\usepackage{amsthm}
\usepackage{footnote}
\usepackage{textcase}
\usepackage{array}
\newcolumntype{P}[1]{>{\centering\arraybackslash}p{#1}}
\usepackage{float} 
\usepackage{multirow}
\usepackage{textcomp}
\usepackage{forest}
\PassOptionsToPackage{hyphens}{url}
\usepackage{url}
\usepackage{physics}
\usepackage{etoolbox}
\usepackage[edges]{forest}
\usepackage[ruled,vlined]{algorithm2e}
\usepackage{siunitx}

\usepackage{subfigure}
\usepackage{physics}
\usepackage{tikz}
\usetikzlibrary{automata, positioning}
\usepackage{nomencl}
\usepackage{float}
\usepackage{tabularx,booktabs}
\renewcommand{\baselinestretch}{0.985}
\usepackage{lipsum}

\usepackage{float}

\begin{document}

%\title{Phasor Domain Dynamic Model of BTB Converter as a Microgrid Building Block (MBB)\\
% delete or comment-out the following line before submission

\title{Dynamic Model of Back-to-Back Converter for System-Level Phasor Simulation \\

\thanks{The Pacific Northwest National Laboratory is operated by the Battelle Memorial Institute for the U.S. Department of Energy under contact DE-AC05-76RL01830.}
}

\author{%%%% author names
    \IEEEauthorblockN{ Hisham Mahmood}% first author
    , \IEEEauthorblockN{ Samrat Acharya}% delete this line if not needed
    , \IEEEauthorblockN{ Francis Tuffner}% delete this line if not needed
        , \IEEEauthorblockN{ Priya Mana}% delete this line if not needed
        , \IEEEauthorblockN{ Alok Kumar Bharati}% delete this line if not needed
    % duplicate the line above as many times as needed to list all authors
    \\%%%% author affiliations
    \IEEEauthorblockA{\textit{Pacific Northwest National Laboratory (PNNL), Richland, WA, USA}}\\% first affiliation

    % duplicate the line above as many times as needed to list all affiliations
    %%%% corresponding author contact details
    \IEEEauthorblockA{(hisham.mahmood, samrat.acharya, francis.tuffner, priya.mana, ak.bharati)@pnnl.gov \vspace{-8mm}}
}

%%%%%%%%%%%%%%%%%%%%%%%%%%%%%%%%%%%%%%%%%%%%%%%%%%%%%%%%%%%%
%      PLEASE DO NOT MAKE ANY CHANGES HERE, USE A PDF COPY
%%%%%%%%%%%%%%%%%%%%%%%%%%%%%%%%%%%%%%%%%%%%%%%%%%%%%%%%%%%%%

\maketitle
\begin{abstract}
   %The power system is rapidly evolving with increasing power electronic components in the grid. The power system operations are also witnessing a significant change with microgrids becoming prominent. Back-to-back converters can be a powerful interface to both grid connected microgrids and networked microgrids. Furthermore, the DC-link of the back-to-back converter can be used to interface and operate hybrid microgrids. However, system-level simulations with detailed models of back-to-back converters may be computationally inefficient. Thus, this study develops a simple phasor domain model that can be easily integrated into  powerflow solvers and facilitate large-scale power system simulations. A phasor domain dynamic model for a back-to-back converter with bidirectional powerflow capability  is developed. The phasor domain dynamic model is validated against an electromagnetic transient (EMT) simulation with detailed switching model.
The power system is expected to evolve rapidly with the increasing deployment of power electronic interface and conditioning systems, microgrids, and hybrid AC/DC grids. %The power system operations are also witnessing a significant change with microgrids becoming more prominent. 
Among power electronic systems, back-to-back (BTB) converters can be a powerful interface to integrate microgrids and networked microgrids. To study the integration of such devices into large power systems, a balance between power electronics model fidelity and system-level computational efficiency is critical. %Furthermore, the DC-link of the back-to-back converter can be used to interface and operate hybrid microgrids. 
%However, system-level simulations with detailed models of back-to-back converters may be computationally inefficient. 
% However, in system-level dynamic simulations for synchronous machine dominated system with BTB converters, a detailed electromagnetic sinusoidal models is not warranted as these simulations are focused on electromechanical transients.
%Dynamic simulations of power systems with detailed electromagnetic sinusoidal model of converters are computationally expensive. However, system-level simulations of power systems with significant synchronous machines do not require detailed converter models as the simulations are focused on electromechanical transients.
 In system-level simulations of bulk power systems dominated by synchronous generators, detailed electromagnetic  models of back-to-back converters may be unnecessary and also computationally inefficient. This paper focuses on developing a simple phasor model for back-to-back converters that can be easily integrated into powerflow solvers to facilitate large-scale power system simulations. 
%large system-level studies with transmission and distribution system solvers
%A phasor domain dynamic model for a back-to-back converter with bidirectional powerflow capability  is developed. 
The model is implemented using C$^{++}$ language and integrated into GridLAB-D, an open source software for distribution systems studies,  as a potential new capability. %The updated version of GridLAB-D with this capability will be publicly available in the near future.  
The GridLAB-D phasor domain model is validated against the electromagnetic transient (EMT) simulation of the detailed switching model. Simulation results show that the phasor model successfully captures the dominant dynamics of the converter with significantly shorter simulation elapsed time.
   
\end{abstract}

\begin{IEEEkeywords}
   Back-to-Back Converter, Phasor Model, Microgrid Building Blocks (MBB), GridLAB-D
\end{IEEEkeywords}

% \blfootnote{This paper is accepted for publication in IEEE PESGM 2024. The complete copyright version will be available on IEEE Xplore when the conference proceedings are published.}
% \vspace{-10pt}
\section{Introduction}
%Back-to-back or BTB converters have been explored substantively in the recent decade for microgrid applications.
%Power systems are evolving rapidly with increasing distributed generation that has been studied in detail in various integrated transmission and distribution studies \cite {akbcosim, akbfreq, akbsmtd}. 

Power systems are evolving rapidly with increasing deployment of power electronics interfaced energy resources, microgrids, and hybrid AC/DC system. 
%The increased penetration of distributed and utility-scale renewables is also leading to the popularity of grid-connected or utility-connected microgrids. 
Among these technologies, the microgrid concept appears to be as one of the promising solutions for the operational and planning challenges facing this grid transformation. To enable fast and reliable deployment of microgrids a concept called microgrid building block (MBB) has been proposed~\cite{mbb_main}. 
%Fast deployment of microgrids and standardized modular approaches to microgrid development is becoming a need to maintain resiliency in the grid that is often adversely impacted by extreme weather events. 
%One of the promising approaches to this modular microgrid is a concept called microgrid building block (MBB). 
MBB is an idea that provides a high-level concept for standardizing and combining power electronic interfaces, communication and control blocks to form and operate microgrids.

The U.S. Department of Energy (DOE) has invested in developing the MBB-based microgrid concept further to standardize the various blocks needed to enable fast deployment of microgrids to provide energy security and energy resilience. A key MBB block is the power converter interface block that has a back-to-back (BTB) converter at the point of common coupling for a microgrid.

The last couple of decades have seen active development of the BTB converter technology that is effectively applied to various microgrid applications. Some of the key features offered by BTB converters include disturbance isolation, grid support, bulk grid interconnection and bidirectional power control and conditioning, etc. \cite{mbbuse1, mbbuse2, mbbuse3}. The key applications that overlap with MBB applications \cite{mbb_main} are:
\begin{enumerate}
    \item Fault isolation;
    \item Power conditioning and power flow control;
    \item Smooth islanding and re-synchronizing (anti-islanding) operation;
    \item DC microgrid integration.
\end{enumerate}
% 1.	Fault isolation;

% 2.	Power conditioning and power flow control;

% 3.	Smooth islanding and re-synchronizing (anti-islanding) operation; 

% 4.	DC microgrid integration.

System level simulations for various configurations of microgrids and networked microgrids built based on the MBB concept will need phasor domain models for the various converters. There are well developed and mature models for smart inverters and other power electronic converters\cite{wei,pephasor1,pephsor2}. However, very few of these models have been integrated and incorporated into three-phase unbalanced power system simulators, as in \cite {gfmgfl1, gfmgfl2, gfmgfl3}. % Inverter models with grid following (GFL) and grid forming (GFM) controls have been integrated into large system-level studies with transmission and distribution system solvers \cite {gfmgfl1, gfmgfl2, gfmgfl3}. %For system-level simulations, it is important to integrate the dynamic models of these converters into similar timescale simulations. This requires phasor domain dynamic models of the power electronic converters that are necessary to enable MBB concept, i.e., the BTB converter.

%%%%%%%%%%%%%%%%%%%%%%%%%%%%%%%%%%%%%%%%%%%%%%%%%%%%%%%%%%%%%%%%%%
%GridLAB-D \cite{chassin2008gridlab} is an open source distribution system solver with several power electronics converters’ dynamic models in the phasor domain, which allows for accurate three-phase unbalanced dynamic simulations at the system-level for distribution system simulations, microgrid studies, and T\&D co-simulations \cite{akbcosim, akbfreq, akbsmtd}.
%%%%%%%%%%%% There is discontinuity with the next paragraph 
%One of the unique aspects of the MBB-based microgrids is the use of BTB converters as power interface for grid connected microgrids and networked microgrids.  

Existing literature has presented detailed EMT simulations for applications of BTB converters in improving the power grid operations \cite{mbbuse1, mbbuse2, mbb1,dclinkcontrol2,conv1}. However, for scalable system-level analysis, % and evaluation of the secondary control schemes for microgrid or power system applications, 
it is important to develop models for the BTB converter that can be used in electromechanical time scale simulations, i.e. phasor domain dynamic models. %The existing models are focused around the control of the BTB converters for specific grid operations \cite{mbb1,dclinkcontrol2,conv1}. 

%However, the fundamental and simple converter-level dynamic models in phasor domain that can be integrated into distribution system solvers are not present in the literature. %The model presented in this paper is developed to be easily integrated into a system-level phasor simulation.

This paper describes the details of a phasor domain dynamic model for a BTB converter that %has a dedicated control for the controlling the DC-link in the BTB converter and another control for controlling the output voltage.
can be integrated into T\&D phasor co-simulation platforms \cite {akbsmtd} where a DC-link could be used to isolate two networks. The $C^{++}$ source code of the phasor model is integrated into GridLAB-D software to validate the model, and to add it as a new capability for this software. GridLAB-D \cite{chassin2008gridlab} is an open source distribution system solver with several power electronics converters’ dynamic models in the phasor domain, and can be used for distribution system, microgrid, and T\&D co-simulations \cite{akbcosim, akbfreq, akbsmtd}. However, GridLAB-D currently does not have the capability of simulating a BTB converter. An updated version of GridLAB-D with this capability, based on this work, will be publicly available in the near future.

% \begin{table*}[H]
% \centering
% \begin{tabular}{ |p{0.8\textwidth}| } % 
%  \hline
%  This paper is accepted for publication in IEEE PESGM 2024. The complete copyright version will be available on IEEE Xplore when the conference proceedings are published. \\
%  \hline
% \end{tabular}
% \end{table*}

The BTB converter model is implemented using three objects representing the two converters and the DC-link to enable its integration into two separate networks isolated by a DC-link. The steps involved in integrating such a model with power system solvers are explained along with the interaction between the dynamic solution of the model and that of the system-level.

\begin{table}[H]
\centering
\begin{tabular}{ |p{0.46\textwidth}| } 
 \hline
 This paper is accepted for publication in IEEE PESGM 2024. The complete copyright version will be available on IEEE Xplore when the conference proceedings are published. \\
 \hline
\end{tabular}
\end{table}

 The GridLAB-D phasor domain BTB converter model is validated against the electromagnetic transient (EMT) simulation of the detailed switching model of the BTB converter developed in MATLAB/Simulink.

The rest of the paper is organized as follows. In Section II, the main phasor equations of the BTB converter are introduced. In Section III, the implementation of the phasor model of the BTB converter is discussed. The GridLAB-D simulation results and validation of the implemented model are presented in Section IV, and conclusions and future work are discussed in Section V.  

\begin{figure*}[!t]
    \centering
\includegraphics[width=0.99\textwidth, clip=true, trim= 0mm 1.1mm 0mm 0mm]{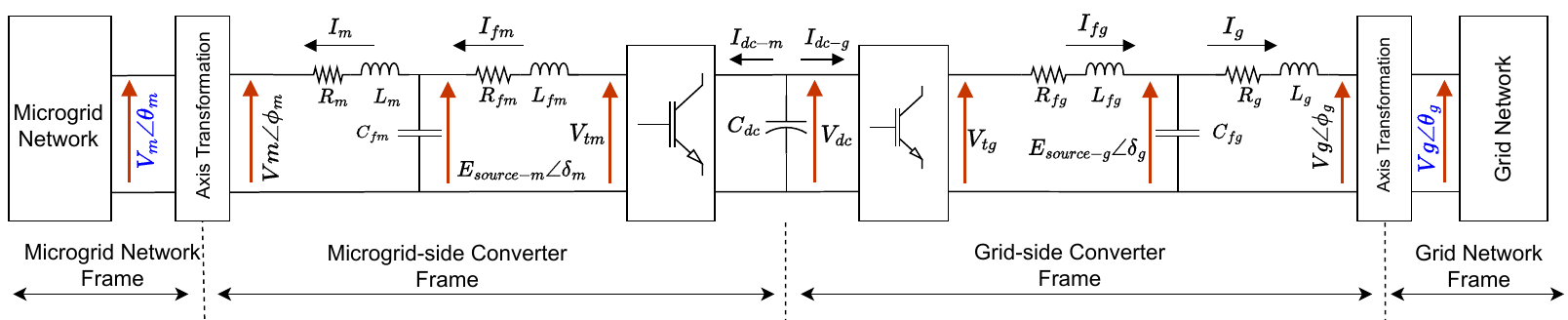}
    \caption{Back-to-back converter connected to a grid and a microgrid network.}
    \label{fig:b2b_config}
\end{figure*}

 \begin{figure}[!t]
     \centering
 \includegraphics[width=0.35\textwidth, clip=true, trim= 0mm 1.1mm 0mm 0mm]{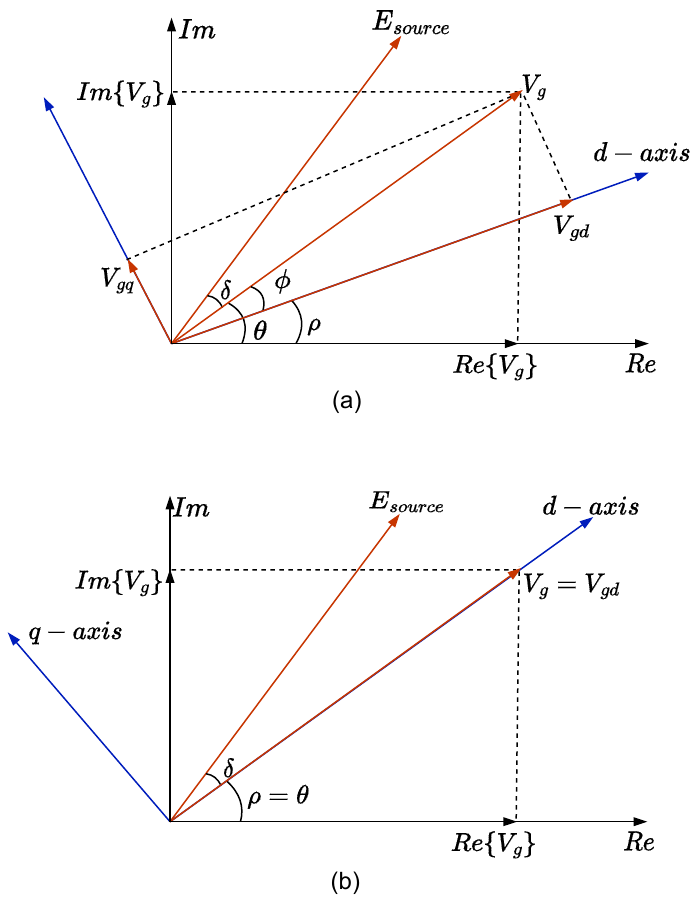}
     \caption{{Network and converter frames, where the dq frame represents the converter side, while the Re/Im frame represent the network side. (a) A general case; (b) the case when the dq frame is aligned with the network terminal voltage using a PLL, i.e. $V_g=V_{gd}$ and $V_q=0$.}}
     \label{fig:Vector_Tr}
 \end{figure}

%\begin{figure}[!t]
%    \centering
%\subfigure[\label{fig:Vdc_2D}]{
%\includegraphics[width=0.48\columnwidth, clip=true, trim= %1mm 86mm 0mm 0mm]{Vector_Tr.pdf}}
%\subfigure[\label{fig:Pg_2D}]{
%\includegraphics[width=0.48\columnwidth, clip=true, trim= %0mm 5mm 0mm 80mm]{Vector_Tr.pdf}}
%\caption{Network and converter frames, where the dq frame represent the converter side, while the Re/Im frame represent the network frame. (a) A general case; (b) the case when the dq frame is aligned with the network terminal voltage using a PLL, i.e. $V_g=V_{gd}$ and $V_q$=0.}
%    \label{fig:Vector_Tr}
%\end{figure}
\vspace{-5mm}
\section{Phasor Equations of the Back-to-Back Converter}
\label{sec:Ph_eqns}
%This section presents a phasor domain model of a back-to-back converter connected to a grid and a microgrid via LCL filters as shown in Fig.\ref{fig:b2b_config}.

Phasor equations of the back-to-back (BTB) converter are introduced in this section to understand the assumptions and simplifications that will be used in Section \ref{sec:model}. A simplified diagram of a typical BTB converter is shown in Fig.\ref{fig:b2b_config}. Subscripts $g$ and $m$ are used for the grid-side converter (GSC) and the microgrid-side converter (MSC), respectively. Fig.\ref{fig:b2b_config} shows an axis transformation block on each side of the BTB converter indicating the transformation from the network frame to the converter frame on each side. These frames are illustrated in Fig.~\ref{fig:Vector_Tr}. Voltages at the points of common coupling (PCC) $V_g\angle{\theta_g}$ and $V_m\angle{\theta_m}$ are provided in the network frames by the power flow solvers on each side during the calculation of the transient response. Phasor equations for the converters are presented below.

%\subsection{Grid-Side Converter Model}
%This section presents the phasor equations for currents and voltages of the grid-side converter (GSC) and their interfaces with DC-link. These equations are used to implement the phasor model of the GSC. 
 The GSC is responsible for regulating the DC-link voltage by controlling its output power in the grid following mode. In this case, the reference for output active power is determined by the DC-link voltage controller. Accordingly, the converter dq frame is synchronized with the grid voltage phasor using a phase-locked-loop (PLL), i.e. $V_g=V_{gd}$ and $V_{gq}$=0 as shown in Fig.~\ref{fig:Vector_Tr}(b), where $V_g\angle{\theta_g}$ is provided by the power flow algorithm, at each time step.

%For this model, the voltage $V_g\angle{\theta}$ at the microgrid point of common coupling (PCC) is considered as an input to the model. This voltage is provided by the powerflow algorithm, at each time step, in the network frame. Converting this voltage into the converter $dq$-frame results in $V_g^d=V_m$ and $V_g^q=0$. 

%Using this voltage and the output currents $I_{g2}^d$ and $I_{g2}^q$ calculated by the DC link controller, the $d$-and $q$-axis voltages at the converter terminal of the GSC LCL filter ($V_{cg}^d$ and $V_{cg}^q)$ are calculated using Kirchhoff's Voltage Law (KVL) as:

The grid current component $I_{gd}$ follows the reference current $I_{gd}^*$ determined by the DC-link controller, while $I_{gq}$ tracks the reference $I_{gq}^*$ which is determined by the voltage controller or the reactive power controller. The variable that plays a major role in integrating the converter into power system solver is the internal voltage (filter Capacitor voltage), which is calculated from Fig.~\ref{fig:b2b_config} using Kirchhoff's Voltage Law (KVL) as:
%\begin{subequations}
%\label{eq:gsc_capacitance_voltage}
%\begin{align}
%    &V_{cg}^d = V_g^d + R_{g2}I_{g2}^d - \omega_g L_{g2}I_{g2}^q,\\
%    &V_{cg}^q = V_g^q + R_{g2}I_{g2}^q- \omega_g L_{g2}I_{g2}^d,
%\end{align}
%\end{subequations}
\begin{subequations}
\label{eq:gsc_capacitance_voltage}
\begin{align}
    &E_{source-gd} = V_{gd} + R_{g}I_{gd} - \omega_g L_{g}I_{gq},\\
    &{E_{source-gq}} = V_{gq} + R_{g}I_{gq} + \omega_g L_{g}I_{gd}
\end{align}
\end{subequations}

%\noindent where $V_g^d$ and $V_g^q$ are $d$-and $q$-axis voltage equivalents of the grid-side PCC voltage, $R_{g2}$ and $L_{g2}$ are the resistance and inductance of the LCL filter at the PCC end, $\omega_g = 2\pi f_g$ is the angular frequency $(f_g)$ of the grid, and $I_{g2}^d$ and $I_{g2}^q$ are the $d$-and $q$-axis currents flowing in $R_{g2}$ and $L_{g2}$. 

Similarly the $d$-and $q$-axis current components of the GSC filter current %$I_{gL}$
$I_{fg}$ are calculated as:
%\begin{subequations}
%    \begin{align}
%        &I_{g1}^d = I_{g2}^d- \omega_g C_{g} V_{cg}^q,\\
%        &I_{g1}^q = I_{g2}^q + \omega_g C_{g} V_{cg}^d,
%    \end{align}
%\end{subequations}

\begin{subequations}
    \begin{align}
        &I_{fg-d} = I_{gd} - \omega_g C_{fg} E_{source-gq},\\
        &I_{fg-q} = I_{gq} + \omega_g C_{fg} E_{source-gd},
    \end{align}
\end{subequations}

\noindent where $\omega_g$  is the radial frequency of the grid. Similarly, the $d$-and $q$-axis voltages at the switching terminals of GSC ($V_{tg-d}$ and $V_{tg-q}$) are calculated using the KVL as:
\begin{subequations}
    \begin{align}
    &V_{tg-d} = E_{source-gd} + R_{fg}I_{fg-d} - \omega_g L_{fg}I_{fg-q},\\
    &V_{tg-q} = E_{source-gq} + R_{fg}I_{fg-q} + \omega_g L_{fg}I_{fg-d}.
    \end{align}
\end{subequations}
%where $R_{g1}$ and $L_{g1}$ are the resistance and inductance of the LCL filter at the GSC end. 

Using $V_{tg}$ and $I_{fg}$, and assuming negligible losses in the converter, the power supplied by the DC-link capacitor into the GSC  $(P_{dc-g})$ and the associated current $I_{dc-g}$ are calculated as:
\begin{subequations}
    \begin{align}
    & P_{dc-g} = \frac{3}{2}(V_{tg-d}I_{fg-d} + V_{tg-q}I_{fg-q}),\\ \label{eq:idc_gsc}
    & I_{dc-g} = \frac{P_{dc-g}}{V_{dc}},
    \end{align}
\end{subequations}
where $V_{dc}$ is the DC-link voltage. Equations (1)-(4) are used also for the MSC by replacing the subscript $g$ with $m$. Accordingly, $V_{dc}$ can be calculated using currents exchanged by the converters with the DC-link as:
 \begin{align}
 \label{eq:Vdc}
     %V_{dc} = \frac{1}{C_{dc}}\int_{t}^{t+\Delta t} (-I_{dc-g} -I_{dc-m})
     V_{dc} = \frac{1}{C_{dc}}\int (-I_{dc-g} -I_{dc-m}) dt,
 \end{align}

\noindent where $I_{dc-m}$ is the DC-link current supplied by the capacitor into the MSC.

%%%%%%%%%%%%%%%%%%%%%%%%%%%%%%%%%%%%%%%%%%%%%%%%%%%%%%%

Since the GSC is responsible for regulating the DC link voltage $V_{dc}$, the grid current component $I_{gd}$ is controlled to follow the reference current $I_{gd}^*$ determined by the active power reference $P_g^*$ generated by the DC-link voltage controller. On the other hand, $I_{gq}$ is determined by the reactive power reference $Q_g^*$. Hence, the reference currents are generated by
\begin{subequations}
\label{eq:gsc_control}
    \begin{align}
        & I_{gd}^* = \frac{2}{3}\frac{P_g^{*}}{V_{gd}},\\
        & I_{gq}^* =\frac{-2}{3} \frac{Q_g^{*}}{V_{gd}},
    \end{align}
where, 
\begin{align}
     & P_g^* = K_P (V_{dc} - V_{dc}^{*}) + K_I \int_{}^{} (V_{dc} - V_{dc}^{*}) dt.
\end{align}
\end{subequations}
The reference $Q_g^*$ is either received as a dispatch command or generated using a local grid voltage controller. In phasor domain simulation of bulk power systems, since the electromechanical dynamics dominate the transient response, the current controller dynamics are approximated by a first-order low pass filter of a certain time constant $T_f$. 
\begin{subequations}
\label{eq:I_filter}
    \begin{align}
        & T_f \frac{d I_{gd}}{dt} +I_{gd} = I_{gd}^*,\\
        & T_f \frac{d I_{gq}}{dt} +I_{gq} = I_{gq}^*.
    \end{align}
\end{subequations}

%%%%%%%%%%%%%%%%%%%%%%%%%%%%%%%%%%%%%%%%%%%%%%%%%%%%%%%%%%%

 \begin{figure}[!t]
     \centering
 \includegraphics[width=0.25\textwidth, clip=true, trim= 0mm 1.1mm 0mm 0mm]{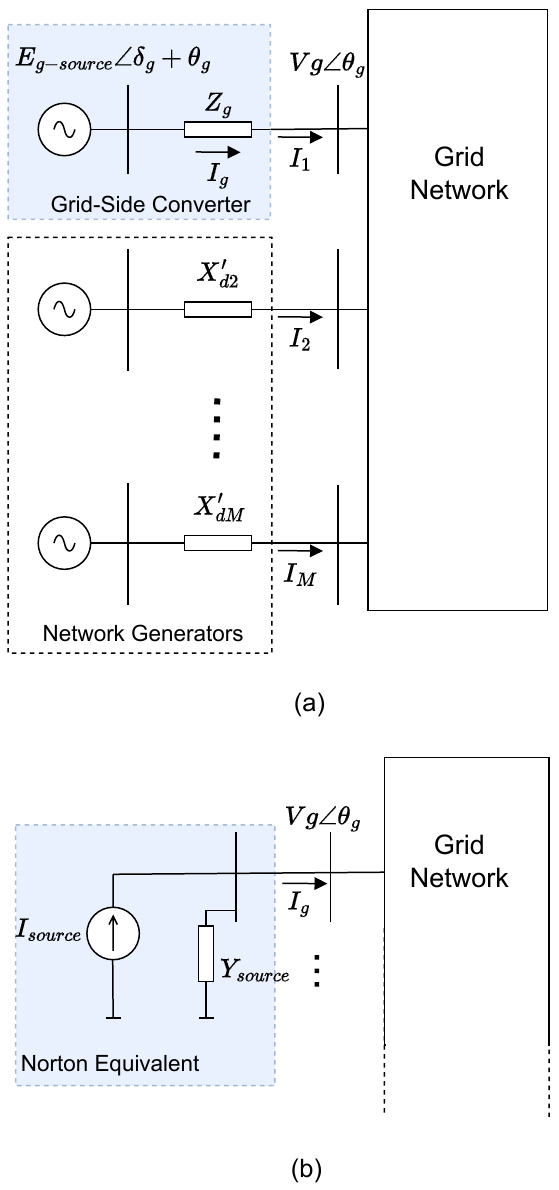}
     \caption{{Thevenin and Norton representation of the GSC and the grid networks. (a) Thevenin equivalent; (b) Norton equivalent}}
     \label{fig:Eqv_Net}
 \end{figure}

%\begin{figure}[!t]
%    \centering
% \includegraphics[width=0.25\textwidth, clip=true, trim= 0mm 1.1mm 0mm 0mm]{Equivalent_Nt.pdf}

%\subfigure[\label{fig:Vdc_2D}]{
%\includegraphics[width=0.48\columnwidth, clip=true, trim= 0mm 85mm 0mm 0mm]{Equivalent_Nt.pdf}}
%\subfigure[\label{fig:Pg_2D}]{
%\includegraphics[width=0.48\columnwidth, clip=true, trim= 0mm 10mm 0mm 125mm]{Equivalent_Nt.pdf}}
%    \caption{Thevenin and Norton representation of the GSC and the grid networks. (a) Thevenin equivalent; (b) Norton equivalent}
%    \label{fig:Eqv_Net}
%\end{figure}

%  \begin{figure}[!t]
% % \vspace{-26pt}
% \centering
% \subfigure[\label{fig:Vdc_2D}]{
% \includegraphics[width=1.05\columnwidth, clip=true, trim= 6.2mm 60mm 10mm 5mm]{Vdc_Pg_2D.eps}}
% \subfigure[\label{fig:Pg_2D}]{
% \includegraphics[width=1.05\columnwidth, clip=true, trim= 6.2mm 8mm 10mm 60mm]{Vdc_Pg_2D.eps}}
% \subfigure[\label{fig:Pmg_2D}]{
% \includegraphics[width=1.05\columnwidth, clip=true, trim= 6.2mm 55mm 10mm 5mm]{P_mg_2D.eps}}
% \caption{Model performance in response to power variations in the exchanged power and power flow reversal.(a) DC-link voltage, (b) Grid side converter power, and (c) Microgrid side converter power.}
% \end{figure}

\begin{figure}[!t]
    \centering
\includegraphics[width=0.48\textwidth, clip=true, trim= 0mm 1.1mm 0mm 0mm]{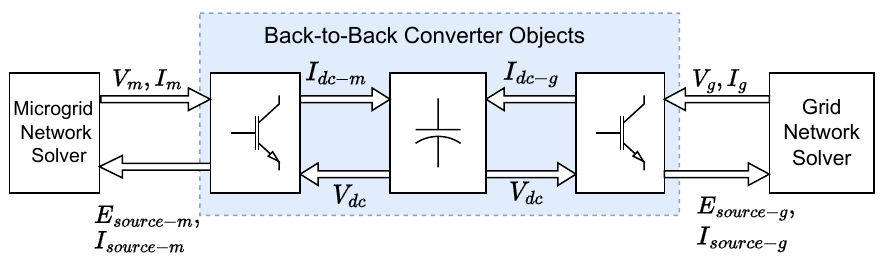}
    \caption{{Three-object structure implementation and data exchange of the back-to-back converter in GridLAB-D.}}
    \label{fig:Conv_Objects}
\end{figure}

\section{Simplified Phasor Model Implementation}
\label{sec:model}

The BTB converter phasor model source code is written in $C^{++}$ and integrated into GridLAB-D software as a new simulation capability. To achieve this integration, the converter on each side of the DC-link is first represented as a voltage source behind an impedance, as shown in Fig.~\ref{fig:Eqv_Net}(a), for the grid side \cite{wei,gfmgfl2}. The voltage behind the impedance is $E_{source}$ which is the voltage at the filter capacitor, whereas the impedance represent the interface reactor and its resistance. Thereafter, the  Norton equivalent circuit is used, as shown in {Fig.~\ref{fig:Eqv_Net}(b)}, to integrate it into the power flow solver of GridLAB-D software.

Because of the unique nature of the BTB converter, where the grid and microgrid networks are isolated and solved separately, the BTB converter is implemented in GridLAB-D using three objects. These objects are the GSC object with the DC-link controller, the MSC object, and the DC-link object, which can exchange data with each other and the network solvers as shown in Fig.~\ref{fig:Conv_Objects}. The focus here will be on the dynamic simulation of the GSC since it is the one considered in charge of the DC-link regulation. Dynamic simulation of the MSC follows similar steps, with more additional flexibility in operating it as a grid forming or a grid following converter. % given that it exchanges information with the DC-link object as shown in Fig.~\ref{fig:Conv_Objects}.

The steps to compute the transient response of the GSC and the DC-link are briefly discussed in the following:

\begin{enumerate}
    \item If a disturbance occurs in the network at time $t$, such as a load change or a fault, the network admittance matrix is updated and loads are calculated as fixed impedances. 
    \item In the network solver, using the initial $E_{source-g}$ or $I_{g-source-g}$ and other generators internal voltages, the bus voltage $V_g$ is calculated at time $t$. Similar calculations are performed for all the generators in the network.

    \item The GSC receives the updated $V_g$ and uses the initial $I_{source-g}$ and Norton equivalent to calculate the initial value for $I_g$ in the network frame.
    
    \item The GSC converts $I_g$ and $V_g$ into the converter frame using the axis transformation in Fig.~\ref{fig:Vector_Tr}(b).

    \item  The GSC uses the transformed values to update $E_{source-g}$ at $t$ using equation (\ref{eq:gsc_capacitance_voltage}).

    \item Ignoring the filter losses, and using the $V_{dc}$ value from the DC-Link object, the power and current at the DC-link can be calculated at time $t$ as:
    \begin{subequations}
    \begin{align}
    & P_{dc-g} = \frac{3}{2}(E_{source-gd}I_{gd} + E_{source-gq}I_{gq}),\\ \label{eq:idc_gsc}
    & I_{dc-g} = \frac{P_{dc-g}}{V_{dc}}.
    \end{align}
\end{subequations}

    \item \textbf{Predictor Iteration:} Execute the first step of  Euler's modified method 
    \begin{enumerate}
        \item Compute the preliminary estimates of the grid current references, $\tilde I_{gd}^*$ and $\tilde I_{gq}^*$, at ($t+\Delta t$) using equation (\ref{eq:gsc_control}).
        \item Compute the preliminary estimates of the grid currents, $\tilde I_{gd}$ and $\tilde I_{gq}$, using equation (\ref{eq:I_filter}) with $\tilde I_{gd}^*$ and $\tilde I_{gq}^*$ as their inputs, respectively.
        \item Calculate $\tilde E_{source-g}$ and $\tilde I_{source-g}$, convert them to the network frame, and provide them to the network solver.
    \end{enumerate}

\item The network solver calculate a preliminary estimate of the bus voltage $\tilde V_g$ at time $t+\Delta t$ 

\item Repeat steps (2-6).

\item \textbf{Corrector Iteration:} Execute the second step of  Euler's modified method to compute the final values of  $I_{gd}$, $I_{gq}$, and $I_{dc-g}$ at $t+\Delta t$.

\item Share $I_{dc-g}$ with the DC-Link object.

\item Calculate the final bus voltage $V_g$ at time $t+\Delta t$ .

\item Set ($t=t+\Delta t$); the DC-link object calculates the DC-link voltage using equation (\ref{eq:Vdc}), and $I_{dc-m}$ and $I_{dc-g}$ values.

\item Stop if $t>T_{stop}$; otherwise, go to Step 1. 
    
\end{enumerate}

\begin{table}[!t]
    \centering
        \caption{Back-to-Back Converter Parameters}
         % \resizebox{0.7\columnwidth}{!}{
\begin{threeparttable}
\small
    \begin{tabular}{|c|c|} 
    \hline
        Parameter & Value  \\
        \hline
        Ph-Ph rms grid voltage, $V_g$ & 208 \unit{V}\\ \hline
        Ph-Ph rms microgrid voltage, $V_m$ & 208 \unit{V}\\ \hline
        Grid-Side Converter kVA, $S_{GSC}$& 50 \unit{kVA}\\ \hline
        Microgrid-Side Converter kVA, $S_{MGSC}$& 50 \unit{kVA}\\ 
        \hline
        $R_{g}$ &  0.001 {\si{\ohm}}\\ \hline
        $L_{g}$ & 0.2 \unit{mH}\\ \hline
        $C_{fg}$  &  50 \unit{\mu F}  \\ \hline
        $R_{fg}$ & 0.001 {\si{\ohm}} \\ \hline
        $L_{fg}$ & 1 \unit{mH}\\ \hline
        $R_{m}$ &  0.001 {\si{\ohm}}\\ \hline
        $L_{m}$ & 0.2 \unit{mH}\\ \hline
        $C_{fm}$  &  50 \unit{\mu F}  \\ \hline
        $R_{fm}$ & 0.001 {\si{\ohm}} \\ \hline
        $L_{fm}$ & 1 \unit{mH}\\ \hline
        $C_{dc}$ & 5000 \unit{\mu F}  \\ \hline
        $V_{dc}^{*}$ & 600 \unit{V} \\ \hline
        Switching frequency (for the detailed model) & 20 \unit{kHz} \\ \hline
        \end{tabular}
    \end{threeparttable}
    % }
    \label{tab:parameters}
\end{table}

\begin{table}[!t]
\centering
{
 \caption{Back-to-Back Converter Control Parameters}
 % \resizebox{0.7\columnwidth}{!}{
\begin{threeparttable}
\small
    \begin{tabular}{|c|c|}
    \hline
        Parameter & Value \\ \hline
        DC-link voltage regulator $K_{P}$&700 $W/V$\\ \hline
        DC-link voltage regulator $K_{I}$& 800 $W/(V.s)$\\ \hline
        Current loop time constant $T_f$&0.005 s\\ \hline
        % DC-link regulator inner loop $K^I_{dcin}$& \\ \hline
        % Power regulator outer loop $K^P_{pout}$&\\ \hline
        % Power regulator outer loop $K^I_{pout}$& \\ \hline
        % Power regulator inner loop $K^P_{pin}$&\\  \hline
        % Power regulator inner loop $K^I_{pin}$& \\ \hline
    \end{tabular}
    \end{threeparttable}
    % }
    \label{tab:control_parameters}
    }
\end{table}

\begin{table}[!t]
    \centering
 \caption{Simulation Time}
 % \resizebox{0.7\columnwidth}{!}{
\begin{threeparttable}
\small
    \begin{tabular}{|c|c|c|}
    \hline
        Simulation Type & Simulation Time & Elapsed Real Time \\ \hline
        EMT & 20 s& 1020 s\\ \hline
        Phasor & 20 s&  9 s\\ \hline
        % DC-link regulator inner loop $K^I_{dcin}$& \\ \hline
        % Power regulator outer loop $K^P_{pout}$&\\ \hline
        % Power regulator outer loop $K^I_{pout}$& \\ \hline
        % Power regulator inner loop $K^P_{pin}$&\\  \hline
        % Power regulator inner loop $K^I_{pin}$& \\ \hline
    \end{tabular}
    \end{threeparttable}
    % }
    \label{tab:sim_time}
    
\end{table}

\section{Validation of Phasor Domain Dynamic Model}
\label{sec:validation}

This section shows the simulation results of the phasor domain model, of the system shown in  Fig.\ref{fig:b2b_config}, implemented in GridLAB-D using the developed objects. The system is also implemented in MATLAB/Simulink to perform EMT simulations. The system and corresponding control parameters are ensured to be the same in both models. The GSC and MSC are connected to ideal voltage sources.

The simulation time steps in GridLAB-D simulations are maintained at 1$\unit{ms}$. The detailed switching model of the BTB converter is simulated at a $1\mu s$ solver time step, and both models are simulated on a computer equipped with Intel(R) Core i7-1185G7 running at 3 GHz, and 16 GB of memory.

The power set-points of MSC are varied to demonstrate the dynamics of the DC-link and the GSC and MSC controllers. %The dynamics of the model implemented in GridLAB-D are compared against the dynamics captured through the EMT simualtions. 
As the BTB model is intended to interface the microgrid at the point of common coupling, two operating scenarios are used for validating the model. The first case is when the microgrid imports power form the grid (grid-side); and, the second case is when the microgrid recers power flow to export power to the grid-side. {Table~\ref{tab:parameters}} shows the grid, microgrid, and BTB converter parameters and {Table~\ref{tab:control_parameters}} shows the key control parameters of the BTB converter.

 The the elapsed real time for the EMT simulation is significantly greater than that required for the phasor simulation as shown in Table \ref{tab:sim_time}.%, which emphasize the value of such phasor models in large system simulations.

\begin{figure}[!t]
    \centering
\includegraphics[width=0.49\textwidth, clip=true, trim= 0mm 1.1mm 0mm 0mm]{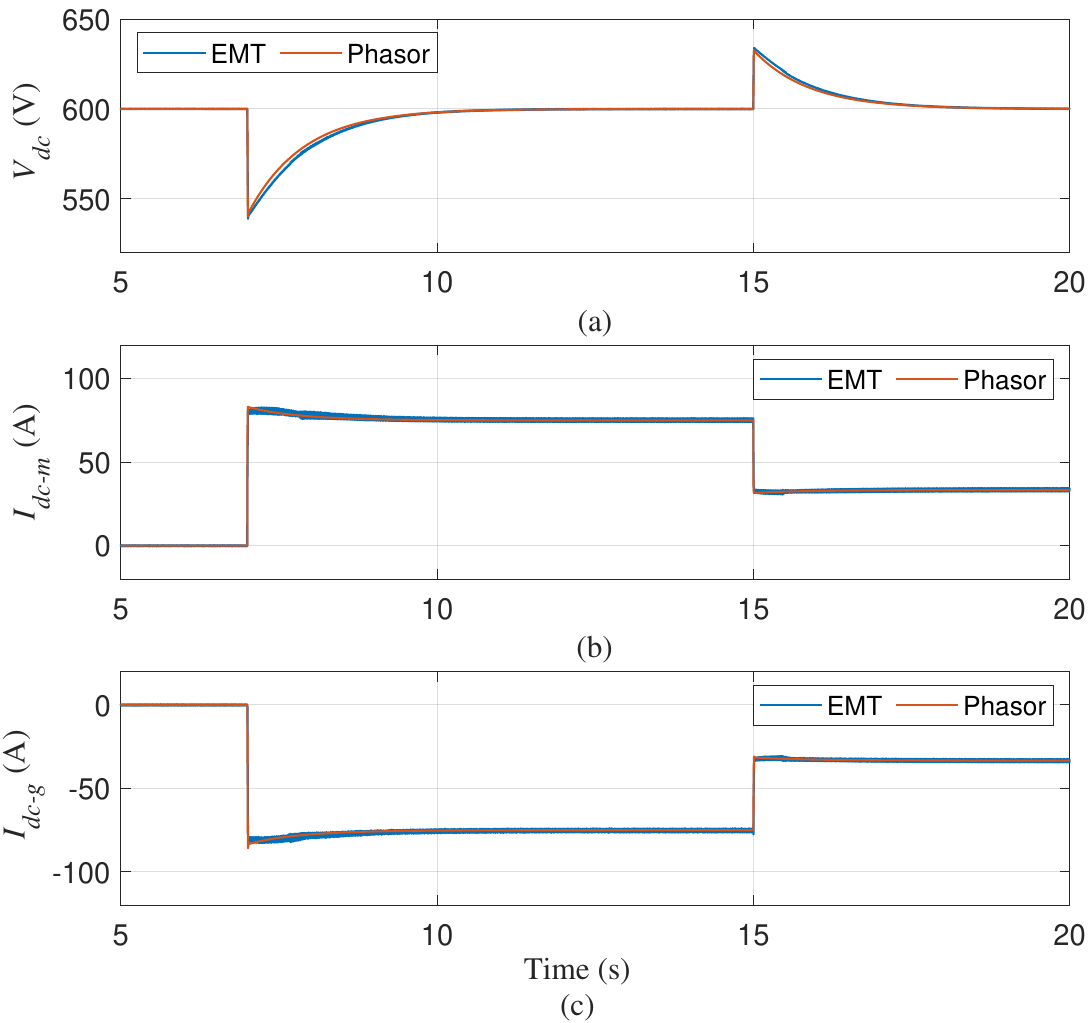}
    \caption{Model performance in response to variations in the power imported by the microgrid. (a) DC-link voltage, (b) DC-link current flowing into the MSC ($I_{dc-m}$), and (c) DC-link current flowing into the GSC ($I_{dc-g}$).}
    \label{fig:1d_flow}
\end{figure}

\vspace{-0.3 cm}

\subsection{Power Flow from Grid to Microgrid}
\label{sec:uni-directional}
This operating scenario is a usual operating scenario for a grid-connected microgrid. %Fig.~\ref{fig:Vdc_1D}-\ref{fig:Pmg_1D} show the dc link voltage and power output at the grid and microgrid converters. 
To focus on the DC-link object performance, Fig.~\ref{fig:1d_flow} shows the DC-link voltage and currents behaviors in response to the changes in the microgrid imported power. In this scenario, the MSC power set point is changed from 0 to 45 kW at t= 7s, and then dropped to 20 kW at t= 15s. To show the dc currents dynamic behavior, the currents in the EMT simulation are filtered with first order {filter} with a time constant of 1 ms.
%Fig.~\ref{fig:Vdc_1D} shows the DC-link dynamics as the power set-point changes (i.e., $P_m = 0$ till 10 seconds, $P_m = 20 kW$ between 10 to 20 seconds, and so on as in Fig.~\ref{fig:Pmg_1D}). 
The results show the transient and the steady-state response of the DC voltage and currents closely match between GridLAB-D and MATLAB/Simulink EMT simulations. %Similar validation results for the power plots on grid-side and microgrid-side are shown in Figs. \ref{fig:Pg_1D} and \ref{fig:Pmg_1D}.
Results also indicate that the proposed three-object implementation of the BTB and the data exchange among objects are integrated successfully into the power flow solver of GridLAB-D software.

\subsection{Bi-Directional Power Flow Between Grid and Microgrid}
\label{sec:bi-directional}
\begin{figure}[!t]
    \centering
\includegraphics[width=0.49\textwidth, clip=true, trim= 0mm 1.1mm 0mm 0mm]{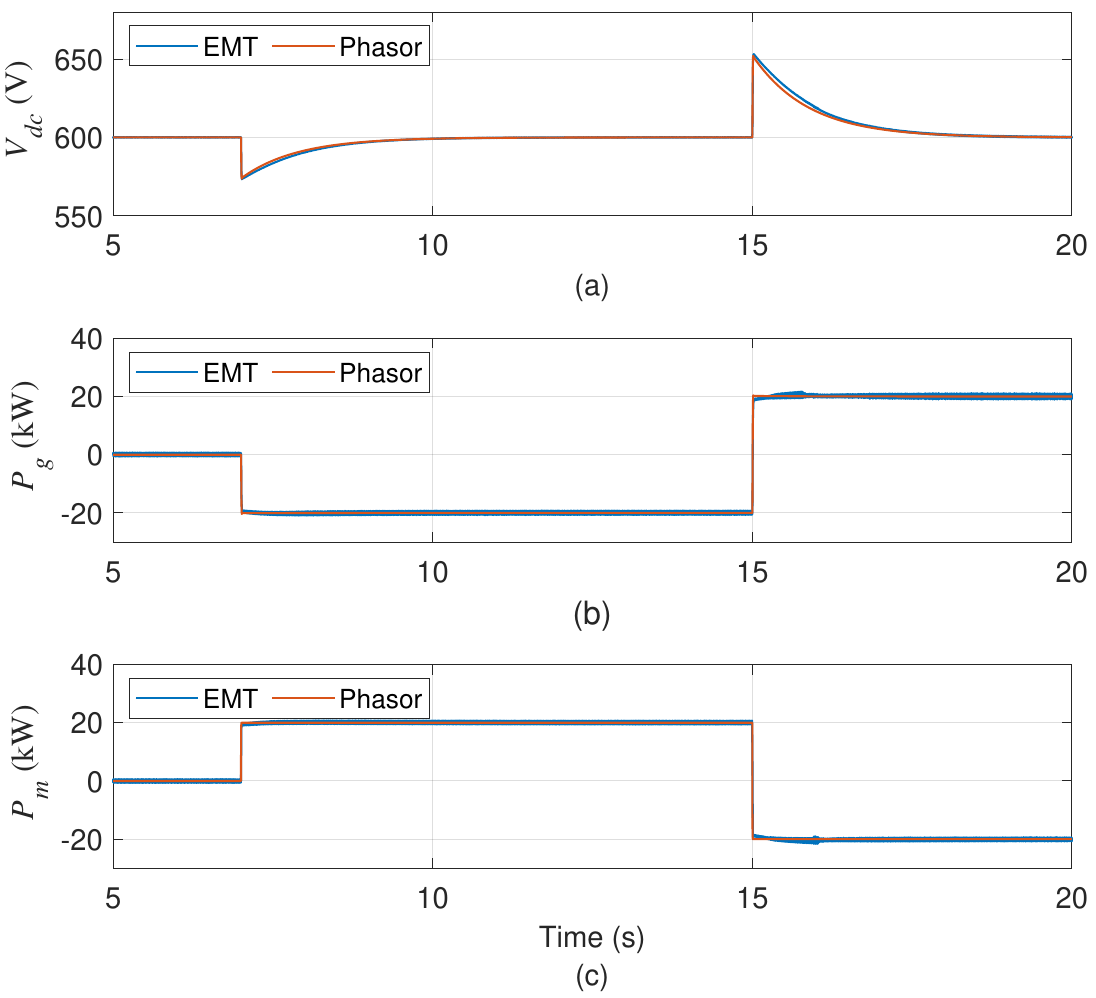}
\caption {Model performance in response to power variations and power flow reversal.(a) DC-link voltage, (b) Grid side converter power ($P_g$), and (c) Microgrid side converter power ($P_m$).}
    \label{fig:2d_flow}
\end{figure}

%This section demonstrates the capability of the BTB converter to exchange power between its connections. 
In this scenario the bi-directional power flow, between grid and microgrid, is examined. Figs.~\ref{fig:2d_flow} show the DC-link voltage and power output at the grid-side and microgrid-side during a power flow direction reversal. 
%In this study, we vary the power set points of MSC (i.e., $P_m = 0$ till 10 seconds, $P_m = 20 kW$ between 10 to 20 seconds, and so on as in Fig.~\ref{fig:Pmg_2D}). 
In this scenario, the MSC power set point ($P_m^*$) is changed from 0 to 20 kW at time $t$=7 s, and then from 20 kW to $-$20 kW at $t$= 15 s. 

%For these power set points, the power output at the GSC for the phasor domain and EMT simulations are shown in Fig.~\ref{fig:Pmg_2D}. Similarly, Fig.~\ref{fig:Pg_2D} shows the power output of GSC for phasor domain and EMT simulations. The DC-link voltage dynamics for the these power exchanges for phasor domain and EMT simulations is shown in Fig.~\ref{fig:Vdc_2D}. The DC-link volatge and power flow between grid and microgrid closely match for the EMT and phasor simluations. 

The responses of the phasor domain model and the EMT model to these set point changes are shown in Fig.~\ref{fig:2d_flow}(b) for the grid-side and in Fig.~\ref{fig:2d_flow}(c) for the microgrid-side. The DC-link volatge and power flow between grid and microgrid closely match for both simulation models.

\section{Conclusions}
%With increasing popularity of microgrids and growing technology advances in power electronics, PCC interfaces like BTB converters are gaining importance to dynamically decouple the microgrid when needed. 
To enable large scale system simulation that focus on electromechanical dynamics, a phasor domain model of BTB converter is developed and validated in this paper. The model is implemented in GridLAB-D software to demonstrate its applicability for system-level dynamic simulations. Preliminary simulations and analysis show that the phasor model captures the dominating dynamics of the detailed switching model. %Simulation scenarios show the bidirectional power flow control capability of the phasor model with regulation of the DC link. 
Simulation results show that the proposed three-object implementation of the BTB and the data exchange among objects are integrated successfully into the power flow solver of GridLAB-D software.
The developed phasor domain model enables easy implementation of BTB converters and can be integrated into other power system software directly or through user defined models. %The steps involved in integrating such a model with power system solvers are explained along with details of interactions of the dynamic solution of the model  with the system-level solution. 
This work will be extended to perform various system-level simulations for different use cases showcasing benefits of interfacing a microgrid using a BTB converter in grid-connected and networked microgrids.

\section*{Acknowledgment}

The authors wish to thank Dan Ton with the U.S. Department of Energy, Office of Electricity for funding this work.  

\bibliographystyle{IEEEtran}
\bibliography{ref}

\begin{thebibliography}{10}
\providecommand{\url}[1]{#1}
\csname url@samestyle\endcsname
\providecommand{\newblock}{\relax}
\providecommand{\bibinfo}[2]{#2}
\providecommand{\BIBentrySTDinterwordspacing}{\spaceskip=0pt\relax}
\providecommand{\BIBentryALTinterwordstretchfactor}{4}
\providecommand{\BIBentryALTinterwordspacing}{\spaceskip=\fontdimen2\font plus
\BIBentryALTinterwordstretchfactor\fontdimen3\font minus \fontdimen4\font\relax}
\providecommand{\BIBforeignlanguage}[2]{{%
\expandafter\ifx\csname l@#1\endcsname\relax
\typeout{** WARNING: IEEEtran.bst: No hyphenation pattern has been}%
\typeout{** loaded for the language `#1'. Using the pattern for}%
\typeout{** the default language instead.}%
\else
\language=\csname l@#1\endcsname
\fi
#2}}
\providecommand{\BIBdecl}{\relax}
\BIBdecl

\bibitem{mbb_main}
\BIBentryALTinterwordspacing
{Building Blocks for Microgrids}. Accessed: June 8, 2023. [Online]. Available: \url{https://www.energy.gov/sites/default/files/2022-09/3-Building\%20Blocks\%20for\%20Microgrids.pdf}
\BIBentrySTDinterwordspacing

\bibitem{mbbuse1}
R.~Majumder, A.~Ghosh, G.~Ledwich, and F.~Zare, ``Power management and power flow control with back-to-back converters in a utility connected microgrid,'' \emph{IEEE Transactions on Power Systems}, vol.~25, no.~2, pp. 821--834, 2010.

\bibitem{mbbuse2}
R.~Majumder, ``A hybrid microgrid with dc connection at back to back converters,'' \emph{IEEE Transactions on Smart Grid}, vol.~5, no.~1, pp. 251--259, 2014.

\bibitem{mbbuse3}
C.-Y. Tang, Y.-F. Chen, Y.-M. Chen, and Y.-R. Chang, ``Dc-link voltage control strategy for three-phase back-to-back active power conditioners,'' \emph{IEEE Transactions on Industrial Electronics}, vol.~62, no.~10, pp. 6306--6316, 2015.

\bibitem{wei}
W.~Du, Y.~Liu, R.~Huang, F.~K. Tuffner, J.~Xie, and Z.~Huang, ``Positive-sequence phasor modeling of droop-controlled, grid-forming inverters with fault current limiting function,'' in \emph{2022 IEEE Power \& Energy Society Innovative Smart Grid Technologies Conference (ISGT)}, 2022, pp. 1--5.

\bibitem{pephasor1}
C.~T. Rim, \emph{Phasor Power Electronics}.\hskip 1em plus 0.5em minus 0.4em\relax Springer, 2016.

\bibitem{pephsor2}
\BIBentryALTinterwordspacing
P.~Santiprapan, K.-L. Areerak, and K.-N. Areerak, ``Mathematical model and control strategy on dq frame for shunt active power filters,'' \emph{International Journal of Electrical and Computer Engineering}, vol.~5, no.~12, pp. 1669 -- 1677, 2011. [Online]. Available: \url{https://publications.waset.org/vol/60}
\BIBentrySTDinterwordspacing

\bibitem{gfmgfl1}
Y.~Li, Y.~Gu, and T.~C. Green, ``Revisiting grid-forming and grid-following inverters: A duality theory,'' \emph{IEEE Transactions on Power Systems}, vol.~37, no.~6, pp. 4541--4554, 2022.

\bibitem{gfmgfl2}
W.~Du, F.~K. Tuffner, K.~P. Schneider, R.~H. Lasseter, J.~Xie, Z.~Chen, and B.~Bhattarai, ``Modeling of grid-forming and grid-following inverters for dynamic simulation of large-scale distribution systems,'' \emph{IEEE Transactions on Power Delivery}, vol.~36, no.~4, pp. 2035--2045, 2021.

\bibitem{gfmgfl3}
A.~Quedan, W.~Wang, D.~Ramasubramanian, E.~Farantatos, and S.~Asgarpoor, ``Dynamic behavior of combined 100\% ibr transmission and distribution networks with grid-forming and grid-following inverters,'' in \emph{2023 IEEE Power \& Energy Society Innovative Smart Grid Technologies Conference (ISGT)}, 2023, pp. 1--5.

\bibitem{mbb1}
Y.-T. Chen, Y.-F. Chen, C.-Y. Tang, Y.-M. Chen, and Y.-R. Chang, ``An active power conditioner with a multi-mode power control strategy for a microgrid,'' in \emph{2013 1st International Future Energy Electronics Conference (IFEEC)}, 2013, pp. 93--97.

\bibitem{dclinkcontrol2}
H.~Akagi and R.~Kitada, ``Control and design of a modular multilevel cascade btb system using bidirectional isolated dc/dc converters,'' \emph{IEEE Transactions on Power Electronics}, vol.~26, no.~9, pp. 2457--2464, 2011.

\bibitem{conv1}
C.~Liu, B.~Wu, N.~R. Zargari, D.~Xu, and J.~Wang, ``A novel three-phase three-leg ac/ac converter using nine igbts,'' \emph{IEEE Transactions on Power Electronics}, vol.~24, no.~5, pp. 1151--1160, 2009.

\bibitem{akbsmtd}
A.~K. Bharati and V.~Ajjarapu, ``\uppercase{SMTD} co-simulation framework with helics for future-grid analysis and synthetic measurement-data generation,'' \emph{IEEE Transactions on Industry Applications}, vol.~58, no.~1, pp. 131--141, 2022.

\bibitem{chassin2008gridlab}
D.~P. Chassin, K.~Schneider, and C.~Gerkensmeyer, ``{GridLAB-D: An open-source power systems modeling and simulation environment},'' in \emph{2008 IEEE/PES Transmission and Distribution Conference and Exposition}.\hskip 1em plus 0.5em minus 0.4em\relax IEEE, 2008, pp. 1--5.

\bibitem{akbcosim}
A.~K. Bharati and V.~Ajjarapu, ``A scalable multi-timescale t\&d co-simulation framework using helics,'' in \emph{2021 IEEE Texas Power and Energy Conference (TPEC)}, 2021, pp. 1--6.

\bibitem{akbfreq}
A.~K. Bharati, V.~Ajjarapu, W.~Du, and Y.~Liu, ``Role of distributed inverter-based-resources in bulk grid primary frequency response through helics based \uppercase{SMTD} co-simulation,'' \emph{IEEE Systems Journal}, vol.~17, no.~1, pp. 1071--1082, 2023.

\end{thebibliography}

\end{document}